\documentclass[12pt,hyperref,showpacs]{revtex4}%
\usepackage{amsfonts}
\usepackage[debug,pdftex]{hyperref}
\usepackage{amsmath}
\usepackage{amssymb}
\usepackage{graphicx}
\usepackage{bibmods}%
\providecommand{\U}[1]{\protect\rule{.1in}{.1in}}
\newtheorem{theorem}{Theorem}
\newtheorem{acknowledgement}[theorem]{Acknowledgement}

\begin{document}
\title{Effect of electromagnetic fields on the creation of scalar particles in a flat
Robertson-Walker space-time}
\author{\textbf{S. Haouat}}
\email{s.haouat@gmail.com}
\affiliation{\textit{LPTh, Department of Physics, University of Jijel, BP 98, Ouled Aissa,
Jijel 18000, Algeria.}}
\author{\textbf{R. Chekireb}}
\affiliation{\textit{LPTh, Department of Physics, University of Jijel, BP 98, Ouled Aissa,
Jijel 18000, Algeria.}}

\begin{abstract}
The influence of electromagnetic fields on the creation of scalar particles
from vacuum in a flat Robertson-Walker space-time is studied. The Klein Gordon
equation with varying electric field and constant magnetic one is solved. The
Bogoliubov transformation method is applied to calculate the pair creation
probability and the number density of created particles. It is shown that the
electric field amplifies the creation of scalar particles while the magnetic
field minimizes it.

\end{abstract}

\pacs{03.65.Pm , 03.70.+k , 04.62.+v }
\maketitle

\section{Introduction}

It is widely known that strong electric field creates particle-antiparticle
pairs from the vacuum. This effect has many important applications in modern
physics from heavy nucleus to black holes \cite{Ruffini}. Several decades ago,
Schwinger studied pair creation effects in the context of gauge invariance and
vacuum polarization \cite{Schwinger}. It has been shown that the vacuum to
vacuum transition amplitude can be expressed through an intermediate effective
action,
\begin{equation}
\mathcal{A}(vac-vac)=\exp(iS_{eff}), \label{1}%
\end{equation}
and the pair creation probability can be extracted from the imaginary part of
this action%
\begin{equation}
\mathcal{P}_{Creat.}=1-\left\vert \mathcal{A}(vac-vac)\right\vert ^{2}%
\simeq2\operatorname{Im}S_{eff}. \label{2}%
\end{equation}
The probability of pair creation from vacuum is calculated in the presence of
some electromagnetic fields \cite{Itzykson,Gitman} and it has been concluded
that electric field produces scalar particles with probability
\begin{equation}
\mathcal{P}_{Creat.}=\frac{e^{2}E^{2}}{8\pi^{3}}\sum_{n=1}^{+\infty}%
\frac{\left(  -1\right)  ^{n-1}}{n^{2}}\exp\left(  -n\pi\frac{m^{2}}%
{eE}\right)  \label{3}%
\end{equation}
while the magnetic field and the plane wave do not create pairs.

However, in spite of the fact that constant magnetic field does not produce
particles, the probability given in (\ref{3}) modifies to be%
\begin{equation}
\mathcal{P}_{Creat.}=\frac{e^{2}EH}{8\pi^{2}}\sum_{n=1}^{+\infty}\frac{\left(
-1\right)  ^{n-1}}{n}{\normalsize \csc}\left(  n\pi\frac{H}{E}\right)
\exp\left(  -n\pi\frac{m^{2}}{eE}\right)  \label{4}%
\end{equation}
when a magnetic field is added to the electric one. Therefore a magnetic field
influences significantly the creation of particles.

Particle-antiparticle pairs may be created also by gravitational fields. This
phenomenon is a prediction of quantum field theory in curved space-time and
its study requires a definition of a vacuum state for the quantum fields
\cite{Birrel,fulling,Mukhanov,Grib,Parker,Winitzki}. However, it is well-known
that in arbitrary curved background, there is no absolute definition of the
vacuum state and the concept of particles is not completely clear. From
physical point of view this is because in quantum theory a particle cannot be
localized to a region smaller than its de Broglie wavelength. When this
wavelength is sufficiently large, the concept of particle becomes unclear. In
expanding universe spontaneous creation of particles occurs because the vacuum
state is unstable - e.g., the vacuum state defined in the remote past differs
from the vacuum state in the remote future. The effect of particle creation
has many applications in contemporary cosmology - e.g., it could have
consequences for early universe cosmology and may play an important role in
the exit from inflationary universe and in the cosmic evolution
\cite{C1,C2,C3,C4,C5,C6,C7}.

The purpose of this paper is to study the effect of electromagnetic fields on
the creation of scalar particles from vacuum in a flat Robertson-Walker
space-time with the use of canonical method based on Bogoliubov relation
between "in" and "out" states. We know that electromagnetic fields are
abundant in the universe during the initial stages of its formation and
certainly these fields have some influences on particle creation. In this
context, electromagnetic fields influence the cosmic evolution directly via
Friedman equations and by their effect on the creation of particles. Motivated
by the fact that there is no electric field in the present stage of the
universe, we consider in this work a varying electric field which vanishes at
$t\rightarrow\infty$. The study of particle creation with stationary electric
field is valid only when the creation of particles is significant for a short
time interval. This is due to the fact that created particles has an inverse
effect on the electric field \cite{Odintsov}. We note that the creation of
scalar particles in electromagnetic and gravitational fields is discussed in
several situations \cite{Odintsov,Villalba1,haouat2,PC1,PC2}.

The paper is organized as follows; At the beginning we introduce a scalar
field propagating in Robertson-Walker space-time and we give a general method
to study the creation of scalar particles in a such space-time. Next we
consider a solvable example with a varying electric field and we give two sets
of exact solutions for the Klein Gordon equation. Then by the use of the
relation between these two sets we determine the probability of pair creation,
the number density of created particles and the vacuum persistence. For a
particular case where the universe behaves like a radiation dominated one we
calculate the total pair production probability from the vacuum to vacuum
transition amplitude and we show how electric field amplifies particle
creation. Finally we consider the combination of electric and magnetic fields.

\section{General formalism for particle creation}

In order to study the phenomenon of particle creation in gravitational fields
we have at our disposal several methods such as the adiabatic method
\cite{Parker1,Parker2,Parker3}, the Hamiltonian diagonalization technique
\cite{Grib1,Grib2,Haro1,Haro2}, the Green function approach
\cite{Gavrilov,Bukhbinder}, the Feynman path integral technique
\cite{Duru,Chitre}, the semiclassical WKB approximation
\cite{Winitzki,Biswas1,Biswas2,Biswas3,Biswas4} as well as the method based on
vacuum to vacuum transition amplitude and Schwinger-like effective action
\cite{Akhme,Kim} and the "in" and "out" states formalism
\cite{Garriga,Villalba2,Villalba3,haouat1} that we shall use in this paper.

To begin let us consider a scalar matter field $\Phi$ with mass $m$ and charge
$e$ subjected to a gravitational field described by the metric $g_{\mu\nu}$
and an electromagnetic field described by the vector $A_{\mu}$. The dynamics
of this system is in general governed by the following Klein Gordon equation%
\begin{equation}
\frac{1}{\sqrt{-g}}\left(  i\partial_{\mu}-eA_{\mu}\right)  \left[  g^{\mu\nu
}\sqrt{-g}\left(  i\partial_{\nu}-eA_{\nu}\right)  \Phi\right]  -\left(
m^{2}+\zeta_{c}R\right)  \Phi=0. \label{5}%
\end{equation}
where $R$ is the Ricci scalar and $\zeta_{c}$ is a numerical parameter (In
conformal coupling $\zeta_{c}=\frac{1}{6}$ ).

We consider, in this work, a flat Robertson-Walker space-time provided with a
metric of the form
\begin{equation}
ds^{2}=dt^{2}-a^{2}\left(  t\right)  \left(  dx^{2}+dy^{2}+dz^{2}\right)
\label{6}%
\end{equation}
This metric can be written as
\begin{equation}
ds^{2}=C\left(  \eta\right)  \left[  d\eta^{2}-dx^{2}-dy^{2}-dz^{2}\right]
\label{7}%
\end{equation}
where $\eta$ is the conformal time $\eta=\int dt/a\left(  t\right)  $ and
$C\left(  \eta\right)  $ is the new scale factor defined by $C\left(
\eta\right)  =\tilde{a}^{2}\left(  \eta\right)  \equiv a^{2}\left[  t\left(
\eta\right)  \right]  .$

We choose to work with conformal time $\eta$ which is convenient to the
present coupling and we consider the gauge $A^{\mu}=$($0$, $0$, $0$,
$A_{z}\left(  \eta\right)  $). If we introduce a new field $\psi\left(
x\right)  $ so that%

\begin{equation}
\Phi\left(  x\right)  =\frac{1}{\sqrt{C\left(  \eta\right)  }}\psi\left(
x\right)  =\frac{1}{\sqrt{C\left(  \eta\right)  }}\chi\left(  \vec{x}\right)
\varphi\left(  \eta\right)  , \label{8}%
\end{equation}
where $\chi\left(  \vec{x}\right)  $ has, in the case of flat space-time, the
form of a plane wave $\chi\left(  \vec{x}\right)  \sim\exp\left(  i\vec
{k}.\vec{r}\right)  ,$ we can obtain the simplified equation
\begin{equation}
\left[  \frac{d^{2}}{d\eta^{2}}+\omega^{2}(\eta)\right]  \varphi\left(
\eta\right)  =0, \label{9}%
\end{equation}
with%
\begin{equation}
\omega^{2}(\eta)=\left[  k_{z}-eA_{z}\left(  \eta\right)  \right]
^{2}+k_{\bot}^{2}+m^{2}C(\eta). \label{10}%
\end{equation}
We assume that the space-time is asymptotically Minkowskian and the potential
$A_{z}\left(  \eta\right)  $ is asymptotically constant when $\eta
\rightarrow\pm\infty$. This choice is suitable for the problem of particles
creation. Since $\omega(\eta)$ in equation (\ref{10}) satisfies the
super-adiabatic condition,
\begin{equation}
\lim\limits_{\eta\rightarrow\pm\infty}\frac{\dot{\omega}}{\omega^{2}}=0
\end{equation}
there is two adiabatic vacuum states and, consequently, the particle
production is well-defined.

In such a case the solutions of the Klein Gordon equation have the following
asymptotic behavior
\begin{align}
\varphi_{in}^{\epsilon}\left(  \eta\right)   &  =\exp\left(  -i\epsilon
\omega_{in}\eta\right) \label{11}\\
\varphi_{out}^{\epsilon}\left(  \eta\right)   &  =\exp\left(  -i\epsilon
\omega_{out}\eta\right)  , \label{12}%
\end{align}
where $\epsilon$ indicates the positive or the negative frequency mode (
$\epsilon=\pm1$) and $\omega_{out}$ and $\omega_{in}$ are given by%
\begin{equation}
\omega_{\substack{in\\out}}=\lim_{\eta\rightarrow\mp\infty}\omega\left(
\eta\right)  \label{13}%
\end{equation}
Let us search for the "\textit{in}" and "\textit{out}" vacuum states. In the
first stage we write the field operator in it's Fourier decomposition
\begin{equation}
\hat{\psi}(\vec{x},\eta)=\frac{1}{\sqrt{2}}\int d^{3}k\left[  \varphi
_{k}^{\ast}\left(  \eta\right)  \chi_{_{k}}\left(  \vec{x}\right)  \hat{a}%
_{k}+\varphi_{k}\left(  \eta\right)  \chi_{_{k}}^{\ast}\left(  \vec{x}\right)
\hat{b}_{k}^{\dag}\right]  \label{14}%
\end{equation}
where, in canonical quantization formalism, the operators $\hat{a}_{k}$ and
$\hat{b}_{k}$ satisfy the following commutation relation%

\begin{equation}
\left[  \hat{a}_{k},\hat{a}_{k^{\prime}}^{\dag}\right]  =\left[  \hat{b}%
_{k},\hat{b}_{k^{\prime}}^{\dag}\right]  =\delta\left(  \vec{k}-\vec
{k}^{\prime}\right)  . \label{15}%
\end{equation}
With the help of the normalization condition%
\begin{equation}
\varphi_{k}^{\ast}\dot{\varphi}_{k}-\varphi_{k}\dot{\varphi}_{k}^{\ast}=2i,
\label{45}%
\end{equation}
we can find without difficulties the following expression of the Hamiltonian
associated with the scalar field system%
\begin{equation}
H=\frac{1}{2}\int d^{3}k\left[  E_{k}\left(  \eta\right)  \left(  \hat{a}%
_{k}\hat{a}_{k}^{\dag}+\hat{b}_{k}^{\dag}\hat{b}_{k}\right)  +F_{k}^{^{\ast}%
}\left(  \eta\right)  \hat{b}_{k}\hat{a}_{k}+F_{k}\left(  \eta\right)  \hat
{a}_{k}^{\dag}\hat{b}_{k}^{\dag}\right]  \label{16}%
\end{equation}
where%
\begin{align}
E_{k}\left(  \eta\right)   &  =\left\vert \dot{\varphi}_{k}\left(
\eta\right)  \right\vert ^{2}+\omega_{k}^{2}\left(  \eta\right)  \left\vert
\varphi_{k}\left(  \eta\right)  \right\vert ^{2}\label{17}\\
F_{k}\left(  \eta\right)   &  =\dot{\varphi}_{k}^{2}\left(  \eta\right)
+\omega_{k}^{2}\left(  \eta\right)  \varphi_{k}^{2}\left(  \eta\right)  .
\label{18}%
\end{align}
Here we remark that $H$ is not diagonal at any time. However, since the mode
functions $\varphi_{k}\left(  \eta\right)  $ have asymptotic behaviors
(\ref{11}) and (\ref{12}) at $\eta\rightarrow\mp\infty$, we can see that
$F_{k}\left(  \eta\right)  =0$ and $H$ becomes diagonal at $\eta\rightarrow
\pm\infty$. In this situation we can define two vacuum states $\left\vert
0_{in}\right\rangle $ and $\left\vert 0_{out}\right\rangle .$ The state
$\left\vert 0_{in}\right\rangle $ is an initial quantum vacuum state in the
remote past with respect to a static observer and $\left\vert 0_{out}%
\right\rangle $ is a final quantum vacuum state in the remote future with
respect to the same observer. This gives some vacuum instability which leads
to particle creation.

Since equation (\ref{9}) is of second order there are only two independent
solutions and all other solutions can be expressed in terms of these two
independent ones. Here we want to find two sets of independent solutions so
that the two functions $\varphi_{in}^{\pm}$ of the first set behave like
positive and negative energy states at $\eta\rightarrow-\infty$ and the two
functions $\varphi_{out}^{\pm}$ of the second set behave like positive and
negative energy states at $\eta\rightarrow+\infty$. The relation between these
two sets or the so-called Bogoliubov transformation is%

\begin{align}
\varphi_{in}^{+}  &  =\alpha\varphi_{out}^{+}+\beta\varphi_{out}^{-}%
\label{19}\\
\varphi_{in}^{-}  &  =\beta^{\ast}\varphi_{out}^{+}+\alpha^{\ast}\varphi
_{out}^{-}, \label{20}%
\end{align}
where the Bogoliubov coefficients $\alpha$ and $\beta$ satisfy the condition
$\left\vert \alpha\right\vert ^{2}-\left\vert \beta\right\vert ^{2}=1.$ The
relation between the creation and annihilation operators is then%

\begin{align}
a_{out}  &  =\alpha\ a_{in}+\beta^{\ast}b_{in}^{\dag}\label{21}\\
b_{out}^{\dag}  &  =\beta\ a_{in}+\alpha^{\ast}b_{in}^{\dag}. \label{22}%
\end{align}
For the process of particle creation the probability amplitude that we want to
calculate is defined by%

\begin{equation}
\mathcal{A}=\left\langle 0_{out}\left\vert a_{out}b_{out}\right\vert
0_{in}\right\rangle . \label{23}%
\end{equation}
Taking into account that\qquad%

\begin{equation}
b_{out}=\frac{1}{\alpha^{\ast}}b_{in}+\frac{\beta^{\ast}}{\alpha^{\ast}%
}a_{out}^{\dag} \label{24}%
\end{equation}
we obtain%

\begin{equation}
\mathcal{A}=\left\langle 0_{out}\left\vert a_{out}b_{out}\right\vert
0_{in}\right\rangle =\frac{\beta^{\ast}}{\alpha^{\ast}}\left\langle
0_{out}\mid0_{in}\right\rangle . \label{25}%
\end{equation}
The probability to create a pair of particles in state $k$ from vacuum is then%

\begin{equation}
\mathcal{P}_{k}=\left\vert \frac{\beta^{\ast}}{\alpha^{\ast}}\right\vert ^{2}.
\label{26}%
\end{equation}
Let $\mathcal{C}_{k}$ to be the probability to have no pair creation in the
state $k.$The quantity $\mathcal{C}_{k}\left(  \mathcal{P}_{k}\right)  ^{n}$
is then the probability to have only $n$ pairs in the state $k.$ We have%

\begin{equation}
\mathcal{C}_{k}+\mathcal{C}_{k}\mathcal{P}_{k}+\mathcal{C}_{k}\left(
\mathcal{P}_{k}\right)  ^{2}+\mathcal{C}_{k}\left(  \mathcal{P}_{k}\right)
^{3}+...=1 \label{27}%
\end{equation}
or simply%

\begin{equation}
\mathcal{C}_{k}=1-\mathcal{P}_{k}. \label{28}%
\end{equation}
Being aware of $\left\vert \frac{\beta^{\ast}}{\alpha^{\ast}}\right\vert
^{2}+\left\vert \frac{1}{\alpha^{\ast}}\right\vert ^{2}=1,$ we can find the
vacuum persistence which reads%
\begin{equation}
\mathcal{C}_{k}=\left\vert \frac{1}{\alpha^{\ast}}\right\vert ^{2}. \label{29}%
\end{equation}
Another important result is the number density of created particles%
\begin{equation}
n\left(  k\right)  =\left\langle 0_{in}\left\vert a_{out}^{+}a_{out}%
\right\vert 0_{in}\right\rangle =\left\vert \beta\right\vert ^{2} \label{30}%
\end{equation}
The general technique for investigating the process of particle creation being
demonstrated, let us give an explicit example where the Klein Gordon equation
admits exact and analytic solutions.

\section{Solvable Model with varying electric field}

Particle creation in Robertson-Walker space-time has been much discussed and
the pair creation probability and the number density of created particles have
been derived for several forms of the scale factor describing different stages
of the evolution of the universe. For the present work we choose for the scale
factor the form
\begin{equation}
C\left(  \eta\right)  =a+b\tanh\left(  \lambda\eta\right)  +c\tanh^{2}\left(
\lambda\eta\right)  \label{31}%
\end{equation}
where $a$, $b$ and $c$ are positive parameters. We can see that this form is
the generalization of various particular cases found in literature; When
$\allowbreak c=0,$ we have a cosmological model with $C\left(  \eta\right)
=a+b\tanh\left(  \lambda\eta\right)  $ which has been widely studied
\cite{Bernard,Setare,Pascoal}. With a particular choice of parameters $a$, $b$
and $c$ we get some models discussed in \cite{Shapoor1,Shapoor2}. In addition,
this universe becomes a radiation dominated one when $a=b=0,$ $c=\frac
{a_{0}^{4}}{4\lambda^{2}}$ and $\lambda\rightarrow0$. We can also make
connection with a Milne universe (i.e. $a\left(  t\right)  =a_{1}t.$) when
$c=0$, $\lambda=a_{1}$, $b=a=\frac{a_{1}^{2}}{2\varepsilon}$ by making the
change $\eta\rightarrow\eta+\frac{\ln\varepsilon}{2\lambda}$and taking the
limit $\varepsilon\longrightarrow0.$

We choose for the varying electric field the gauge%
\begin{equation}
A_{\mu}=\frac{E_{0}}{\lambda}\tanh\lambda\eta~\delta_{\mu3}, \label{32}%
\end{equation}
which describes the following electric field%
\begin{equation}
\vec{E}=\frac{1}{C\left(  \eta\right)  }\frac{E_{0}}{\cosh^{2}\lambda\eta}%
\vec{u}_{z}.
\end{equation}
This field becomes the so-called Sauter field in the case of Minkowski
space-time (i.e. $C\left(  \eta\right)  =1$ and $\eta=t$) \cite{Kim3}.

In such a case $\omega_{out}$ and $\omega_{in}$ are given by%
\begin{align}
\omega_{in} &  =\sqrt{k^{2}+m^{2}\left(  a+\bar{c}-\bar{b}\right)  }%
\label{33}\\
\omega_{out} &  =\sqrt{k^{2}+m^{2}\left(  a+\bar{c}+\bar{b}\right)
}.\label{34}%
\end{align}
and the simplified Klein Gordon becomes
\begin{equation}
\left(  \frac{d^{2}}{d\eta^{2}}+k^{2}+m^{2}\left(  a+\bar{b}\tanh\lambda
\eta+\bar{c}\tanh^{2}\lambda\eta\right)  \right)  \varphi=0\label{35}%
\end{equation}
with%
\begin{equation}
\left.
\begin{array}
[c]{c}%
\bar{b}=b-\frac{2ek_{z}E_{0}}{\lambda m^{2}}\\
\bar{c}=c+\left(  \frac{eE_{0}}{\lambda m}\right)  ^{2}%
\end{array}
\right.  \label{36}%
\end{equation}
Now in order to solve equation (\ref{35}) we make the change $\eta
\rightarrow\xi,$ where%

\begin{equation}
\xi=\frac{1+\tanh\left(  \lambda\eta\right)  }{2}. \label{37}%
\end{equation}
The resulting equation that takes the form%

\begin{align}
&  \left[  \frac{\partial^{2}}{\partial\xi^{2}}+\left(  \frac{1}{\xi}-\frac
{1}{1-\xi}\right)  \frac{\partial}{\partial\xi}+\left(  \frac{\omega_{in}^{2}%
}{4\lambda^{2}}\frac{1}{\xi}-\frac{m^{2}c}{\lambda^{2}}\right.  \right.
\nonumber\\
&  \left.  \left.  \left.  +\frac{\omega_{out}^{2}}{4\lambda^{2}}\frac
{1}{\left(  1-\xi\right)  }\right)  \frac{1}{\xi\left(  1-\xi\right)
}\right]  \tilde{\varphi}\left(  \xi\right)  =0\right.  \label{38}%
\end{align}
is a Riemann type equation \cite{Grad}
\begin{align}
&  \left[  \frac{\partial^{2}}{\partial\xi^{2}}+\left(  \frac{1-\alpha
_{1}-\alpha_{1}^{\prime}}{\xi}-\frac{1-\alpha_{3}-\alpha_{3}^{\prime}}{1-\xi
}\right)  \frac{\partial}{\partial\xi}+\right. \nonumber\\
&  \left.  \left.  \left(  \frac{\alpha_{1}\alpha_{1}^{\prime}}{\xi}%
-\alpha_{2}\alpha_{2}^{\prime}+\frac{\alpha_{3}\alpha_{3}^{\prime}}{1-\xi
}\right)  \frac{1}{\xi\left(  1-\xi\right)  }\right]  \tilde{\varphi}\left(
\xi\right)  =0\right.  \label{39}%
\end{align}
where%

\begin{equation}%
\begin{array}
[c]{l}%
\alpha_{1}=-\alpha_{1}^{\prime}=i\frac{\omega_{in}}{2\lambda}\\
\alpha_{3}=-\alpha_{3}^{\prime}=i\frac{\omega_{in}}{2\lambda}\\
\alpha_{2}=1-\alpha_{2}^{\prime}=\frac{1}{2}+\sqrt{\frac{m^{2}\bar{c}}%
{\lambda^{2}}-\frac{1}{4}},
\end{array}
\label{40}%
\end{equation}
with the condition $\alpha_{1}+\alpha_{1}^{\prime}+\alpha_{2}+\alpha
_{2}^{\prime}+\alpha_{3}+\alpha_{3}^{\prime}=1.$

Following \cite{Grad} we can find for equation (\ref{39}) several sets of
solutions that can be written in terms of hypergeometric functions. Taking
into account the behavior of positive and negative energy states we can
classify our two sets as follows; for the "in" states we have%

\begin{align}
\tilde{\varphi}_{in}^{+}\left(  \xi\right)   &  =\frac{1}{\sqrt{2\omega_{in}}%
}\xi^{-i\frac{\omega_{in}}{2\lambda}}(1-\xi)^{i\frac{\omega_{out}}{2\lambda}%
}\nonumber\\
&  F\left(  \frac{1}{2}+i\frac{\omega_{-}}{\lambda}+i\delta,\frac{1}{2}%
+i\frac{\omega_{-}}{\lambda}-i\delta;1-i\frac{\omega_{in}}{\lambda}%
;\xi\right)  \label{41}%
\end{align}
and%
\begin{align}
\tilde{\varphi}_{in}^{-}\left(  \xi\right)   &  =\frac{1}{\sqrt{2\omega_{in}}%
}\xi^{i\frac{\omega_{in}}{2\lambda}}(1-\xi)^{-i\frac{\omega_{out}}{2\lambda}%
}\nonumber\\
&  F\left(  \frac{1}{2}-i\frac{\omega_{-}}{\lambda}+i\delta,\frac{1}{2}%
-i\frac{\omega_{-}}{\lambda}-i\delta;1+i\frac{\omega_{in}}{\lambda}%
;\xi\right)  , \label{42}%
\end{align}
with%
\begin{equation}
\omega_{\pm}=\frac{\omega_{out}\pm\omega_{in}}{2} \label{43}%
\end{equation}
and%
\begin{equation}
\delta=\frac{1}{2}\sqrt{\frac{4m^{2}\bar{c}}{\lambda^{2}}-1}. \label{44}%
\end{equation}
The factors $\left(  2\omega_{in}\right)  ^{-1/2}$ and $\left(  2\omega
_{out}\right)  ^{-1/2}$ are determined by the use of the normalization
condition (\ref{45}) which explains the conservation of the Klein Gordon
particle current.

For the "out" states we have%

\begin{align}
\tilde{\varphi}_{out}^{+}\left(  \xi\right)   &  =\frac{1}{\sqrt{2\omega
_{out}}}\xi^{-i\frac{\omega_{in}}{2\lambda}}(1-\xi)^{i\frac{\omega_{out}%
}{2\lambda}}\nonumber\\
&  F\left(  \frac{1}{2}+i\frac{\omega_{-}}{\lambda}+i\delta,\frac{1}{2}%
+i\frac{\omega_{-}}{\lambda}-i\delta;1+i\frac{\omega_{out}}{\lambda}%
;1-\xi\right)  \label{46}%
\end{align}
and%
\begin{align}
\tilde{\varphi}_{out}^{-}\left(  \xi\right)   &  =\frac{1}{\sqrt{2\omega
_{out}}}\xi^{i\frac{\omega_{in}}{2\lambda}}(1-\xi)^{-i\frac{\omega_{out}%
}{2\lambda}}\nonumber\\
&  F\left(  \frac{1}{2}-i\frac{\omega_{-}}{\lambda}+i\delta,\frac{1}{2}%
-i\frac{\omega_{-}}{\lambda}-i\delta;1-i\frac{\omega_{out}}{\lambda}%
;1-\xi\right)  . \label{47}%
\end{align}
Let us, now use the relation between "in" and "out" solutions to determine the
probability of pair creation and the number density of created particles. By
the use of the relation between hypergeometric functions \cite{Grad}%
\begin{align}
F\left(  u,v;w;\xi\right)   &  =\frac{\Gamma\left(  w\right)  \Gamma\left(
w-v-u\right)  }{\Gamma\left(  w-u\right)  \Gamma\left(  w-v\right)  }F\left(
u,v;u+v-w+1;1-\xi\right)  ~~\nonumber\\
&  +\left(  1-\xi\right)  ^{w-u-v}\frac{\Gamma\left(  \gamma\right)
\Gamma\left(  u+v-w\right)  }{\Gamma\left(  u\right)  \Gamma\left(  v\right)
}\nonumber\\
&  F\left(  w-u,w-v;w-v-u+1;1-\xi\right)  \label{48}%
\end{align}
and the property%
\begin{equation}
F\left(  u,v;w;\xi\right)  =\left(  1-\xi\right)  ^{w-u-v}F\left(
w-u,w-v;w;\xi\right)  , \label{49}%
\end{equation}
we obtain%
\begin{equation}
\alpha=\sqrt{\frac{\omega_{out}}{\omega_{in}}}\frac{\Gamma\left(
1-i\frac{\omega_{in}}{\lambda}\right)  \Gamma\left(  -i\frac{\omega_{out}%
}{\lambda}\right)  }{\Gamma\left(  \frac{1}{2}-i\frac{\omega_{+}}{\lambda
}-i\delta\right)  \Gamma\left(  \frac{1}{2}-i\frac{\omega_{+}}{\lambda
}+i\delta\right)  } \label{50}%
\end{equation}
and%
\begin{equation}
\beta=\sqrt{\frac{\omega_{out}}{\omega_{in}}}\frac{\Gamma\left(
1-i\frac{\omega_{in}}{\lambda}\right)  \Gamma\left(  i\frac{\omega_{out}%
}{\lambda}\right)  }{\Gamma\left(  \frac{1}{2}+i\frac{\omega_{-}}{\lambda
}-i\delta\right)  \Gamma\left(  \frac{1}{2}+i\frac{\omega_{-}}{\lambda
}+i\delta\right)  }. \label{51}%
\end{equation}
The probability to create one pair of particles from vacuum is then%

\begin{equation}
\mathcal{P}_{k}=\left\vert \frac{\Gamma\left(  \frac{1}{2}-i\frac{\omega_{+}%
}{\lambda}-i\delta\right)  \Gamma\left(  \frac{1}{2}-i\frac{\omega_{+}%
}{\lambda}+i\delta\right)  }{\Gamma\left(  \frac{1}{2}+i\frac{\omega_{-}%
}{\lambda}-i\delta\right)  \Gamma\left(  \frac{1}{2}+i\frac{\omega_{-}%
}{\lambda}+i\delta\right)  }\right\vert ^{2}. \label{52}%
\end{equation}
Using the following properties of the Gamma functions \cite{Grad}%
\begin{equation}
\Gamma(z+1)=z\Gamma(z), \label{53}%
\end{equation}%
\begin{equation}
\left\vert \Gamma(ix)\right\vert ^{2}=\frac{\pi}{x\sinh\pi x} \label{54}%
\end{equation}
and%
\begin{equation}
\left\vert \Gamma(\frac{1}{2}+ix)\right\vert ^{2}=\frac{\pi}{\cosh\pi x}
\label{55}%
\end{equation}
we arrive at
\begin{equation}
\mathcal{P}_{k}=\frac{\cosh\left(  2\pi\frac{\omega-}{\lambda}\right)
+\cosh\left(  \pi\sqrt{\frac{4m^{2}\bar{c}}{\lambda^{2}}-1}\right)  }%
{\cosh\left(  2\pi\frac{\omega+}{\lambda}\right)  +\cosh\left(  \pi\sqrt
{\frac{4m^{2}\bar{c}}{\lambda^{2}}-1}\right)  }. \label{56}%
\end{equation}
For the vacuum persistence we obtain%
\begin{equation}
\mathcal{C}_{k}=\frac{\cosh\left(  2\pi\frac{\omega+}{\lambda}\right)
-\cosh\left(  2\pi\frac{\omega-}{\lambda}\right)  }{\cosh\left(  2\pi
\frac{\omega+}{\lambda}\right)  +\cosh\left(  \pi\sqrt{\frac{4m^{2}\bar{c}%
}{\lambda^{2}}-1}\right)  }. \label{57}%
\end{equation}
A simple calculation gives for the number density of created particles%

\begin{equation}
n\left(  k\right)  =\frac{\cosh\left(  2\pi\frac{\omega_{-}}{\lambda}\right)
+\cosh\left(  \pi\sqrt{\frac{4m^{2}\bar{c}}{\lambda^{2}}-1}\right)  }%
{\cosh\left(  2\pi\frac{\omega+}{\lambda}\right)  -\cosh\left(  2\pi
\frac{\omega-}{\lambda}\right)  }. \label{58}%
\end{equation}
Here we note that the number density of created particles can be written as%
\begin{equation}
n\left(  k\right)  =\frac{1}{\left\vert \frac{\alpha}{\beta}\right\vert
^{2}-1}, \label{59}%
\end{equation}
and for large frequencies $n\left(  k\right)  $ becomes%
\begin{equation}
n\left(  k\right)  =\frac{1}{\exp\left(  \frac{2\pi}{\lambda}\omega
_{in}\right)  -1} \label{60}%
\end{equation}
which is a thermal Bose-Einstein distribution.

Let us note that when\textbf{ }$b=c=0$\ and $a=1$ and by taking the
limit\textbf{ }$\lambda\rightarrow0$, we obtain the well-known result
associated with the constant electric field in Minkowski space-time%
\begin{equation}
\mathcal{P}_{k}=\frac{\exp\left(  -\pi\frac{k_{\perp}^{2}+m^{2}}{eE_{0}%
}\right)  }{1+\exp\left(  -\pi\frac{k_{\perp}^{2}+m^{2}}{eE_{0}}\right)  },
\label{61}%
\end{equation}
with $k_{\perp}^{2}=k_{x}^{2}+k_{y}^{2}.$

\section{Particular case}

Now we consider a cosmological model with a scale factor of the form%
\begin{equation}
C\left(  \eta\right)  =a+\frac{a_{_{0}}^{4}}{4}\eta^{2} \label{62}%
\end{equation}
which describes a radiation dominated like universe $a\left(  t\right)  \sim
a_{_{0}}\sqrt{t}.$ It is obvious that this situation can be obtained by
considering the particular case when $b=0$, $c=\frac{a_{_{0}}^{4}}%
{4\lambda^{2}}$\ and by taking the limit $\lambda\rightarrow0$. Here the role
of the parameter $a$ is to check the correctness of our results by making
comparison to the case of Minkowski space-time when $a=1$ and $a_{0}=0$. To
consider the particle creation in pure radiation dominated universe we have to
put $a=0$.

It is easy to show that when the scale factor is given by equation (\ref{62}),
the probability $\mathcal{P}_{k}$ can be written in the form
\begin{equation}
\mathcal{P}_{k}=\frac{\sigma}{1+\sigma}, \label{63}%
\end{equation}
where%
\begin{equation}
\sigma=\exp\left[  \allowbreak-2\pi\left(  \frac{k_{\bot}^{2}+am^{2}}%
{\sqrt{m^{2}a_{_{0}}^{4}+4e^{2}E_{0}^{2}}}\allowbreak+\frac{m^{2}a_{0}%
^{4}k_{z}^{2}}{\left(  m^{2}a_{0}^{4}+4e^{2}E_{0}^{2}\right)  ^{\frac{3}{2}}%
}\allowbreak\right)  \right]  . \label{64}%
\end{equation}
The vacuum to vacuum transition probability is then%

\begin{equation}
\exp\left(  -2\operatorname{Im}S_{eff}\right)  =\prod_{k}\mathcal{C}_{k}%
=\prod_{k}\exp\left[  -\ln\left(  1+\sigma\right)  \right]  \label{65}%
\end{equation}
and consequently%

\begin{equation}
2\operatorname{Im}S_{eff}=\sum_{k}\ln\left(  1+\sigma\right)  . \label{66}%
\end{equation}
Expanding the quantity $\ln(1+\sigma)$ and replacing the summation over $k$ by
$\int\frac{d^{3}k}{\left(  2\pi\right)  ^{3}},$ we get%

\begin{equation}
2\operatorname{Im}S_{eff}=\int\frac{d^{3}k}{\left(  2\pi\right)  ^{3}}%
\ \sum_{n=1}\frac{\left(  -1\right)  ^{n+1}}{n}\exp\left[  -2n\pi\left(
\frac{k_{\bot}^{2}+am^{2}}{\sqrt{m^{2}a_{_{0}}^{4}+4e^{2}E_{0}^{2}}}%
+\frac{m^{2}a_{0}^{4}k_{z}^{2}\allowbreak}{\left(  m^{2}a_{0}^{4}+4e^{2}%
E_{0}^{2}\right)  ^{\frac{3}{2}}}\allowbreak\right)  \right]  . \label{67}%
\end{equation}
By doing integration over $k_{x}$ and $k_{y}$ we obtain%
\begin{align}
2\operatorname{Im}S_{eff}  &  =\frac{\sqrt{m^{2}a_{_{0}}^{4}+4e^{2}E_{0}^{2}}%
}{2\left(  2\pi\right)  ^{3}}\ \sum_{n=1}\frac{\left(  -1\right)  ^{n+1}%
}{n^{2}}\exp\left[  -2n\pi\left(  \frac{am^{2}}{\sqrt{m^{2}a_{_{0}}^{4}%
+4e^{2}E_{0}^{2}}}\allowbreak\right)  \right] \nonumber\\
&  \int dk_{z}\exp\left[  -2n\pi\left(  \frac{m^{2}a_{0}^{4}}{\left(
m^{2}a_{0}^{4}+4e^{2}E_{0}^{2}\right)  ^{\frac{3}{2}}}\allowbreak\right)
k_{z}^{2}\right]  . \label{68}%
\end{align}
For the integration over $k_{z}$ we can use the following property%
\begin{equation}
dk_{z}=\frac{m^{2}a_{_{0}}^{4}+4e^{2}E_{0}^{2}}{4eE_{0}}d\eta\label{69}%
\end{equation}
to write $2\operatorname{Im}S_{eff}$ in the form%
\begin{equation}
2\operatorname{Im}S_{eff}=\int d\eta\Gamma\left(  \eta\right)  \label{70}%
\end{equation}
where the\ particle creation probability per unit of time $\Gamma\left(
\eta\right)  $ is given by
\begin{align}
\Gamma\left(  \eta\right)   &  =\frac{\left(  2e\mathcal{E}\right)  ^{3}%
}{8\left(  2\pi\right)  ^{3}eE_{0}}\ \sum_{n=1}\frac{\left(  -1\right)
^{n+1}}{n^{2}}\exp\left[  -n\pi\left(  \frac{am^{2}}{e\mathcal{E}}%
\allowbreak\right)  \right] \nonumber\\
&  \exp\left[  -2n\pi\left(  \frac{m^{2}a_{0}^{4}}{16e^{2}E_{0}^{2}%
}\allowbreak2e\mathcal{E}\right)  \eta^{2}\right]  . \label{71}%
\end{align}
with%
\begin{equation}
4e^{2}\mathcal{E}^{2}=m^{2}a_{_{0}}^{4}+4e^{2}E_{0}^{2}. \label{72}%
\end{equation}
Here we note that we obtain the Schwinger result by setting $a=1$ and
$a_{_{0}}=0.$

For $a_{_{0}}\not =0,$ by doing integration over conformal time $\eta,$ we get
a Schwinger-like series
\begin{equation}
2\operatorname{Im}S_{eff}=\frac{1}{4\pi^{3}}\frac{\left(  e\mathcal{E}\right)
^{\frac{5}{2}}}{ma_{0}^{2}}\ \sum_{n=1}\frac{\left(  -1\right)  ^{n+1}%
}{n^{\frac{5}{2}}}\exp\left[  -n\pi\frac{m^{2}a}{e\mathcal{E}}\allowbreak
\right]  \label{73}%
\end{equation}
Now if we consider a pure radiation dominated universe (i.e. $a=0$) we can see
that the electric field amplifies the gravitational particle creation by the
following factor%
\begin{equation}
\gamma=\frac{2\operatorname{Im}S_{eff}}{2\operatorname{Im}S_{eff}\left(
E_{0}=0\right)  }=\left(  1+4\frac{e^{2}E_{0}^{2}}{m^{2}a_{_{0}}^{4}}\right)
^{\frac{5}{4}}\ .\label{74}%
\end{equation}
which is, for strong field, of order $\left(  eE_{0}/ma_{_{0}}^{2}\right)
^{\frac{5}{2}}.$

For the number density of created particles we have%
\begin{equation}
n\left(  k\right)  =\exp\left[  -2\pi\left(  \frac{k_{\bot}^{2}}{\sqrt
{m^{2}a_{_{0}}^{4}+4e^{2}E_{0}^{2}}}+\frac{m^{2}a_{0}^{4}k_{z}^{2}\allowbreak
}{\left(  m^{2}a_{0}^{4}+4e^{2}E_{0}^{2}\right)  ^{\frac{3}{2}}}%
\allowbreak\right)  \right]  .\label{75}%
\end{equation}
The total number of created particles can be written then in the form%
\begin{equation}
N_{T}=\frac{\left(  m^{2}a_{_{0}}^{4}+4e^{2}E_{0}^{2}\right)  ^{\frac{3}{2}}%
}{8\left(  2\pi\right)  ^{3}eE_{0}}\allowbreak\int d\eta\exp\left[
-\pi\left(  \frac{m^{2}a_{0}^{4}}{8e^{2}E_{0}^{2}}\allowbreak\sqrt
{m^{2}a_{_{0}}^{4}+4e^{2}E_{0}^{2}}\right)  \eta^{2}\right]  .\label{76}%
\end{equation}
Consequently we have%
\begin{equation}
\frac{dN}{d\eta}=n\left(  \eta\right)  =\frac{\left(  m^{2}a_{_{0}}^{4}%
+4e^{2}E_{0}^{2}\right)  ^{\frac{3}{2}}}{8\left(  2\pi\right)  ^{3}eE_{0}%
}\allowbreak\exp\left[  -\pi\left(  \frac{m^{2}a_{0}^{4}}{8e^{2}E_{0}^{2}%
}\allowbreak\sqrt{m^{2}a_{_{0}}^{4}+4e^{2}E_{0}^{2}}\right)  \eta^{2}\right]
.\label{77}%
\end{equation}
By doing integration over $\eta$ we get%
\begin{equation}
N_{T}=\frac{\left(  m^{2}a_{_{0}}^{4}+4e^{2}E_{0}^{2}\right)  ^{\frac{5}{4}}%
}{\sqrt{8}\left(  2\pi\right)  ^{3}ma_{0}^{2}}.\allowbreak\label{78}%
\end{equation}
We see that the factor $\gamma$ can be obtained also from $N_{T}$
\begin{equation}
\frac{N_{T}}{N_{T}\left(  E_{0}=0\right)  }=\gamma,\label{79}%
\end{equation}
This effect seems to be important for light particles. However, this is not
true. Since $n\left(  \eta\right)  $ in equation (\ref{77}) is Gaussian with
respect to $\eta$ we find that particle creation is significant in the time
interval%
\begin{equation}
\Delta\eta=\frac{1}{\sqrt{2\pi}}\frac{4eE_{0}}{\allowbreak ma_{0}^{2}}\left(
m^{2}a_{_{0}}^{4}+4e^{2}E_{0}^{2}\right)  ^{-\frac{1}{4}}.
\end{equation}
In radiation dominated universe, electromagnetic backgrounds can be considered
quasi-stationary only for short time $\Delta\eta<<1.$ This gives
\begin{equation}
ma_{0}^{2}>>\sqrt{eE_{0}}.
\end{equation}
The effect of electric field is then more important when the mass of created
particles verifies the condition$\sqrt{eE_{0}}<<\allowbreak ma_{0}^{2}%
<<eE_{0}$. Thus, electric field predominantly produces heavy particles.
Furthermore, it is possible to create super-heavy particles with the mass of
the Grand Unification scale in the early universe by strong electric field.
This may have many important cosmological consequences. For light particles
the effect of the electric field on particle creation is negligible although
the factor $\gamma$ becomes large when $m$ is small. This is explained by the
fact that, when $m$ decreases $N_{T}\left(  E_{0}=0\right)  $ decreases so
that $N_{T}=\gamma N_{T}\left(  E_{0}=0\right)  $ remains negligible.

\section{Effect of magnetic field}

Having studied the phenomenon of particle creation in the presence of an
electric field, let us now consider the superposition of an electric field and
a magnetic one to investigate the influence of the magnetic fields on the
creation of scalar particles. For this aim we choose the gauge%
\begin{equation}
A_{\mu}=\frac{E_{0}}{\lambda}\tanh\left(  \lambda\eta\right)  ~\delta_{\mu
3}-Hx~\delta_{\mu2} \label{80}%
\end{equation}
which leads to the following Klein Gordon equation%

\begin{equation}
\left[  \frac{\partial^{2}}{\partial\eta^{2}}-\frac{\partial^{2}}{\partial
x^{2}}+\left(  i\frac{\partial}{\partial y}+eHx\right)  ^{2}+\left(
i\frac{\partial}{\partial z}-\frac{eE_{0}}{\lambda}\tanh\lambda\eta\right)
^{2}+m^{2}C\left(  \eta\right)  \right]  \psi\left(  \eta,\vec{x}\right)  =0.
\label{81}%
\end{equation}
To solve this equation we decompose $\psi\left(  \eta,\vec{x}\right)  $ as%
\begin{equation}
\psi\left(  \eta,\vec{x}\right)  =\varphi(\eta)g(x)\exp\left(  -ik_{y}%
y-ik_{z}z\right)  , \label{82}%
\end{equation}
where the functions $\varphi(\eta)$ and $g(x)$ obey respectively%
\begin{equation}
\left[  \frac{d^{2}}{d\eta^{2}}+\left(  k_{z}-\frac{eE_{0}}{\lambda}%
\tanh\lambda\eta\right)  ^{2}+m^{2}C\left(  \eta\right)  +\kappa\right]
\varphi(\eta)=0 \label{83}%
\end{equation}
and%
\begin{equation}
\left[  -\frac{d^{2}}{dx^{2}}+\left(  k_{y}+eHx\right)  ^{2}\right]
g(x)=\kappa g(x) \label{84}%
\end{equation}
where $\kappa$ is a constant resulting from the separation of variables. It is
clear that by making the change $x\rightarrow x-\frac{k_{y}}{eH},$ equation
(\ref{84}) becomes similar to the wave equation associated with the harmonic
oscillator, where the solution is given by%

\begin{equation}
g\left(  x\right)  =\left(  \frac{1}{2^{l}l!\sqrt{\pi}}\right)  ^{\frac{1}{2}%
}\left(  \frac{eH}{4}\right)  ^{\frac{1}{2}}\exp\left[  -\frac{eH}{2}\left(
x+\frac{k_{y}}{eH}\right)  ^{2}\right]  \ \mathcal{H}_{l}\left[  \sqrt
{eH}\left(  x+\frac{k_{y}}{eH}\right)  \right]  \label{85}%
\end{equation}
and
\begin{equation}
\kappa=eH\left(  2l+1\right)  . \label{86}%
\end{equation}
Here, $l$ is an integer and $\mathcal{H}_{l}\left(  x\right)  $ is the Hermit
polynomial. For the function $\varphi(\eta)$ we have
\begin{equation}
\left[  \frac{d^{2}}{d\eta^{2}}+k_{z}^{2}+m^{2}\left(  a^{\prime}+\bar{b}%
\tanh\lambda\eta+\bar{c}\tanh^{2}\lambda\eta\right)  \right]  \varphi(\eta)=0
\label{87}%
\end{equation}
where%
\begin{equation}
a^{\prime}=a+\frac{eH}{m^{2}}\left(  2l+1\right)  \label{88}%
\end{equation}
The later equation is similar to (\ref{35}) with the change $a\rightarrow
a^{\prime}.$ Then with the same steps as in section (III) we can obtain the
following results%
\begin{equation}
\mathcal{P}_{k,l}=\frac{\cosh\left(  2\pi\frac{\omega_{-}^{\prime}}{\lambda
}\right)  +\cosh\left(  2\pi\delta\right)  }{\cosh\left(  2\pi\frac{\omega
_{+}^{\prime}}{\lambda}\right)  +\cosh\left(  2\pi\delta\right)  }, \label{89}%
\end{equation}
where%
\begin{align}
\omega_{in}^{\prime}  &  =\sqrt{k_{z}^{2}+m^{2}\left(  a^{\prime}+\bar{c}%
-\bar{b}\right)  }\label{90}\\
\omega_{out}^{\prime}  &  =\sqrt{k_{z}^{2}+m^{2}\left(  a^{\prime}+\bar
{c}+\bar{b}\right)  }. \label{91}%
\end{align}
Let us remark here that when \textbf{ }$a=1$, $b=c=0$\textbf{\ }and\textbf{
}$\lambda\rightarrow0$ we obtain the probability of pair creation in Minkowski
space-time with electric and magnetic fields
\begin{equation}
\mathcal{P}_{k,l}=\frac{\exp\left(  -\pi\frac{m^{2}+eH\left(  2l+1\right)
}{eE_{0}}\right)  }{1+\exp\left(  -\pi\frac{m^{2}+eH\left(  2l+1\right)
}{eE_{0}}\right)  }. \label{92}%
\end{equation}
Like in the previous section when $b=0$, $c=\frac{c_{_{0}}}{\lambda^{2}}%
$\textbf{\ }and\textbf{ }$\lambda\rightarrow0$ we can get easily%

\begin{align}
2\operatorname{Im}S_{eff}  &  =\int\frac{dk_{y}dk_{z}}{\left(  2\pi\right)
^{2}}\sum_{l}\ \sum_{n=1}\frac{\left(  -1\right)  ^{n+1}}{n}\times\nonumber\\
&  \exp\left[  -n\pi\left(  \frac{eH\left(  2l+1\right)  +am^{2}}%
{e\mathcal{E}}+\frac{m^{2}a_{0}^{4}k_{z}^{2}\allowbreak}{4\left(
e\mathcal{E}\right)  ^{3}}\allowbreak\right)  \right]  . \label{93}%
\end{align}
Taking into account that%
\begin{equation}
dk_{y}=eHdx \label{94}%
\end{equation}
we can write $2\operatorname{Im}S_{eff}$ as follows%
\begin{equation}
2\operatorname{Im}S_{eff}=\int dxd\eta\Gamma\left(  x,\eta\right)  ,
\label{95}%
\end{equation}
where $\Gamma\left(  x,\eta\right)  $ is the pair creation probability per
unit of time per unit of volume. It is given by%
\begin{align}
\Gamma\left(  x,\eta\right)   &  =\frac{eH}{\left(  2\pi\right)  ^{2}}%
\frac{e^{2}\mathcal{E}^{2}}{eE_{0}}\sum_{n=1}\frac{\left(  -1\right)  ^{n+1}%
}{n}\sum_{l}\ \exp\left[  -n\pi\frac{H}{\mathcal{E}}\left(  2l+1\right)
\right]  \times\nonumber\\
&  \exp\left[  -n\pi\left(  \frac{am^{2}}{e\mathcal{E}}+\frac{m^{2}a_{0}%
^{4}\allowbreak e\mathcal{E}}{4e^{2}E_{0}^{2}}\eta^{2}\right)  \right]  .
\label{96}%
\end{align}
Note here that $\Gamma\left(  x,\eta\right)  $ does not depend on $x$
\ because de universe is homogeneous\ and the magnetic field too.

By summing over $l$%
\begin{equation}
\sum_{l=0}\ \exp\left[  -n\pi\frac{H}{\mathcal{E}}\left(  2l+1\right)
\right]  =\frac{1}{2\sinh\left(  n\pi\frac{H}{\mathcal{E}}\right)  }
\label{97}%
\end{equation}
we get the Schwinger-like series%
\begin{equation}
\Gamma\left(  \eta\right)  =\frac{1}{\left(  2\pi\right)  ^{3}}\frac{\left(
e\mathcal{E}\right)  ^{3}}{eE_{0}}\sum_{n=1}\frac{\left(  -1\right)  ^{n+1}%
}{n^{2}}F_{n}\left(  H\right)  \exp\left[  -n\pi\left(  \frac{am^{2}%
}{e\mathcal{E}}+\frac{m^{2}a_{0}^{4}\allowbreak e\mathcal{E}}{4e^{2}E_{0}^{2}%
}\eta^{2}\right)  \right]  , \label{98}%
\end{equation}
where the factor $F_{n}\left(  H\right)  $ that describes the effect of the
magnetic field is given by%
\begin{equation}
F_{n}\left(  H\right)  =\frac{n\pi\frac{H}{\mathcal{E}}}{\sinh\left(
n\pi\frac{H}{\mathcal{E}}\right)  }. \label{99}%
\end{equation}
We remark that $0<F_{n}\left(  H\right)  \leq1$. This means that the magnetic
field minimizes the creation of scalar particles.

For the number density we obtain%

\begin{equation}
N_{T}\left(  H,E_{0}\right)  =\int d\eta\frac{\pi\frac{H}{\mathcal{E}}}%
{\sinh\left(  \pi\frac{H}{\mathcal{E}}\right)  }\frac{4e^{2}\mathcal{E}^{3}%
}{4\left(  2\pi\right)  ^{3}eE_{0}}\exp\left[  -\pi\left(  \frac{am^{2}%
}{e\mathcal{E}}+\frac{m^{2}a_{0}^{4}\allowbreak e\mathcal{E}}{4e^{2}E_{0}^{2}%
}\eta^{2}\right)  \right]  .\label{100}%
\end{equation}
In the radiation era (i.e. $a=0$), we have
\begin{equation}
n\left(  \eta\right)  =\frac{\pi\frac{H}{\mathcal{E}}}{\sinh\left(  \pi
\frac{H}{\mathcal{E}}\right)  }\frac{4e^{2}\mathcal{E}^{3}}{4\left(
2\pi\right)  ^{3}eE_{0}}\exp\left[  -\pi\frac{m^{2}\allowbreak e\mathcal{E}%
}{e^{2}E_{0}^{2}}C\left(  \eta\right)  \right]  .\label{101a}%
\end{equation}
This result is in complete agreement with equation (5.8) in \cite{PC1}.

Now by doing integration over $\eta$ we obtain the expression%
\begin{equation}
N_{T}=\left(  1+\frac{4e^{2}E_{0}^{2}}{m^{2}a_{_{0}}^{4}}\right)  ^{\frac
{5}{4}}\frac{\pi\frac{H}{\mathcal{E}}}{\sinh\left(  \pi\frac{H}{\mathcal{E}%
}\right)  }\frac{\left(  ma_{_{0}}^{2}\right)  ^{\frac{3}{2}}}{2\sqrt
{2}\left(  2\pi\right)  ^{3}}, \label{101}%
\end{equation}
which reduces in the case of a pure magnetic field to%

\begin{equation}
N_{T}\left(  H,E_{0}=0\right)  =\frac{\sqrt{ma_{_{0}}^{2}}}{8\sqrt{2}\pi^{2}%
}\frac{eH}{\sinh\left(  2\pi\frac{eH}{ma_{_{0}}^{2}}\right)  }. \label{102}%
\end{equation}
Thus the magnetic field minimizes the gravitational particle creation by the
factor%
\begin{equation}
\gamma^{\prime}=\frac{2\pi\frac{eH}{ma_{_{0}}^{2}}}{\sinh\left(  2\pi\frac
{eH}{ma_{_{0}}^{2}}\right)  }<1.
\end{equation}
In the presence of both fields the amplification factor will be given by%
\begin{equation}
\gamma^{\prime\prime}=\frac{\pi Y\left(  1+X^{2}\right)  ^{\frac{3}{4}}}%
{\sinh\left(  \pi\frac{Y}{\sqrt{1+X^{2}}}\right)  },
\end{equation}
where $X=\frac{2eE_{0}}{ma_{_{0}}^{2}}$ and $Y=\frac{2eH}{ma_{_{0}}^{2}}$. In
this case $\gamma^{\prime\prime}$ may be less than $1.$ This depends on the
values of $E_{0}$ and $H$ . When $H\sim E_{0}>>ma_{_{0}}^{2},$ we can see that
$\gamma^{\prime\prime}>>1$ and the creation of super-heavy particles increases.

In addition, it is well-known that pure gravitational fields do not\ create
massless particles with conformal coupling. Unlike results of reference
\cite{PC2} this remains true even if electromagnetic fields are present. In
effect from equation (\ref{101}) we can see that $N_{T}=0$ when $m=0$.

\section{Conclusion}

In this paper we have studied the effect of electromagnetic fields on the
creation of scalar particles in a Robertson-Walker space-time by considering
the canonical method based on Bogoliubov transformation. $\allowbreak$We have
given two sets of exact solutions for the Klein Gordon field equation with
varying electric field and we have used these solutions to calculate the
probability of pair creation and the number density of created particles.

Then we have discussed a particular cosmological model that behaves like
radiation dominated universe where we have calculated the vacuum to vacuum
transition probability and we have extracted the nonvanishing imaginary term
of the effective action that means that created particles are real and not
virtual ones. We have considered also the combination of varying electric
field and constant homogenous magnetic field.

The essential result is that strong electric field amplifies gravitational
particle creation by a factor of order $\left(  eE_{0}/ma_{_{0}}^{2}\right)
^{\frac{5}{2}}.$ This conclusion is in agreement with the result of
\cite{Odintsov}. We have shown also that the magnetic field minimizes the
particle creation like in case of Minkowski space-time with pure
electromagnetic fields. Then the effect of electric field is more important
than the magnetic field one vis-a-vis the process of cosmological scalar
particle creation.

It is obvious that the inclusion of magnetic field may be done by making the
change $k_{\bot}^{2}\rightarrow\left(  2n+1\right)  $ $eH.$This explains why
the magnetic field minimizes the particles creation - e.g., since particles
prefer to be created in lower energy state \cite{Gitman} and the minimum of
$k_{\bot}^{2}$ is $0$ while the minimum of $\left(  2n+1\right)  $ $eH$ is
$eH$.

In addition, the creation of massless particles with conformal coupling is
impossible even if electromagnetic fields are present.

\begin{acknowledgement}
The authors wish to thank the referees for their precious comments which
greatly improved the paper. The work of S Haouat is partially supported by
Algerian Ministry of High Education and Scientific Research and\ ANDRU under
the PNR project: COSMOGR (code: 8/u18/976, contract n$%
%TCIMACRO{\U{b0}}%
%BeginExpansion
{{}^\circ}%
%EndExpansion
$ 28/04)
\end{acknowledgement}


\begin{thebibliography}{99}                                                                                               %


\bibitem {Ruffini}R. Ruffini, G. Vereshchagin, S-S. Xue, Phys. Rep.
\textbf{487} (2010)

\bibitem {Schwinger}J. Schwinger, Phys. Rev. \textbf{82,} 664 (1951)

\bibitem {Itzykson}E. Bresin and C. Itzykson, Phys. Rev. D \textbf{2,
}1191\textbf{ }(1970).

\bibitem {Gitman}E.S. Fradkin, D.M. Gitman and S.M. Shvartsman, Quantum
Electrodynamics with Unstable Vacuum (Springer-Verlag, Berlin 1991)

\bibitem {Birrel}N.D.Birrell and P. C. W.Davies, Quantum Fields in Curved
Space (Cambridge Univ. Press, Cambridge1982).

\bibitem {fulling}S. A. Fulling, Aspects of Quantum Field Theory in Curved
Space-Time (Cambridge University Press, Cambridge 1985)

\bibitem {Mukhanov}V. F. Mukhanov and S. Winitzki; Introduction to Quantum
Effects in Gravity (Cambridge Univ. Press, Cambridge, 2007)

\bibitem {Grib}A. A.Grib, S. G. Mamayev, and V.M.Mostepanenko,Vacuum Quantum
Effects in Strong Fields ( Friedmann Lab. Publ.,St. Petersburg 1994)

\bibitem {Parker}L. Parker and D. J. Toms Quantum Field Theory in Curved
Space-time: Quantized Fields and Gravity (Cambridge University Press,
Cambridge 2009)

\bibitem {Winitzki}S. Winitzki; Phys. Rev. D \textbf{72,} 104011 (2005)

\bibitem {C1}S. Debnath and A. K. Sanyal; Class. Quantum Grav. \textbf{28},
145015 (2011)

\bibitem {C2}J. A. S. Lima, F. E. Silva and R. C. Santos, Class. Quantum.
Grav. \textbf{25}, 205006 (2008).

\bibitem {C3}V. Sahni and S. Habib, Phys. Rev. Lett. \textbf{81}, 1766 (1998)

\bibitem {C4}L. Kofman: Preheating After Inflation, Lect. Notes Phys.
\textbf{738}, 55 (2008) (Springer-Verlag Berlin Heidelberg 2008)

\bibitem {C5}J. Martin: Inflationary Perturbations: The Cosmological Schwinger
Effect, Lect. Notes Phys.\textbf{738}, 193 (2008) (Springer-Verlag Berlin
Heidelberg 2008)

\bibitem {C6}I. Antoniadis, P. O. Mazur and E.Mottola, New J. Phys.
\textbf{9}, 11 (2007)

\bibitem {C7}J. Chen, P. Wu, H. Yu, Z. Li, Eur. Phys. J. C \textbf{72}, 1861 (2012)

\bibitem {Villalba1}V. M. Villalba, Phys. Rev. D \textbf{60}, 127501 (1999)

\bibitem {haouat2}S. Haouat and R. Chekireb, Int. J. Theor. Phys. \textbf{51},
1704 (2012)

\bibitem {PC1}G.\ Sch\"{a}fer et H.\ Dehnen : J. Phys. A: Math. Gen,
\textbf{13}, 517 (1980).

\bibitem {PC2}S. Moradi; Mod. Phys. Lett. A \textbf{24}, 1129 (2009).

\bibitem {Odintsov}I L Buchbinder and S D Odintsov, Sov. Phys. J. \textbf{25},
385 (1982)

\bibitem {Parker1}L.Parker, Phys. Rev. Lett. \textbf{21}, 562 (1968).

\bibitem {Parker2}L.Parker, Phys.Rev. D \textbf{183}, 1057 (1969).

\bibitem {Parker3}L.Parker, Phys.Rev. D \textbf{3}, 346 (1971).

\bibitem {Grib1}A. A. Grip, S. G.\ Mamayev and V.\ M.\ Mostepanenko : Gen.
Rel. Grav. \textbf{7}, 535 (1976).

\bibitem {Grib2}A. A. Grip, S. G.\ Mamayev, and V.\ M.\ Mostepanenko, J. Phys.
A: Math. Gen \textbf{13,} 2057 (1980)

\bibitem {Haro1}J. Haro and E. Elizald; J.Phys. A: Math. Theor. \textbf{41,}
372003 (2008).

\bibitem {Haro2}J. Haro;J.Phys. A: Math. Theor. \textbf{44} (2011).

\bibitem {Gavrilov}S. Gavrilov, D.M. Gitman, and S.D. Odintsov, Int. J. Mod.
Phys. A \textbf{12}, 4837 (1997)

\bibitem {Bukhbinder}I.L. Bukhbinder, Izv. Vyssh. Uchebn. Zaved. Fiz.
\textbf{7,} 3 (1980).

\bibitem {Duru}I. H. Duru, N.\ \"{U}nal : Phys. Rev. D \textbf{34}, 966 (1986).

\bibitem {Chitre}D. M. Chitre and J. B. Hartle, Phys. Rev D. \textbf{16}, 251 (1977).

\bibitem {Biswas1}S. Biswas, J. Guha and N. G. Sarkar; Class. Quantum Grav.
\textbf{12}, 1591 (1995)

\bibitem {Biswas2}J. Guha, D. Biswas, N. G. Sarkar and S. Biswas; Class.
Quantum Grav. \textbf{12,} 1641 (1995)

\bibitem {Biswas3}S. Biswas, A. Shaw and P. Misra; Gen. Rel. Grav.\textbf{
34}, 665 (2002)

\bibitem {Biswas4}S. Biswas and I. Chowdhury; Int. J. Mod. Phys. D
\textbf{15,} 937 (2006)

\bibitem {Akhme}E. Akhmedov; Mod. Phys. Lett. A \textbf{25,} 2815 (2010)

\bibitem {Kim}S. P. Kim; arXiv:1008.0577v1 [hep-th]

\bibitem {Garriga}J. Garriga; Phys. Rev. D \textbf{49}, 6343 (1994)

\bibitem {Villalba2}V. M. Villalba and W. Greiner; Phys. Rev. D \textbf{65,}
025007 (2001)

\bibitem {Villalba3}V. M.\ Villalba; Progress of Theoretical Physics,
\textbf{90}, 851 (1993).

\bibitem {haouat1}S. Haouat and R. Chekireb, Mod. Phys. Lett. A. \textbf{26},
2653 (2011)

\bibitem {Bernard}C. Bernard, A. Duncun, Ann. Phys. \textbf{107,} 201 (1977).

\bibitem {Setare}M.R. Setare, Int. J. Theor. Phys. \textbf{43}, 2237 (2004).

\bibitem {Pascoal}F. Pascoal, C. Farina, Int. J. Theor. Phys. \textbf{46} (2007)

\bibitem {Shapoor1}S. Moradi; Int. J. Theor. Phys. \textbf{47,} 2808 (2008).

\bibitem {Shapoor2}S. Moradi; Journal of Geometry and Physics \textbf{59,} 173 (2009).

\bibitem {Kim3}S. P. Kim, H. K. Lee and Y. Yoon; Phys. Rev. D \textbf{78,}
105013 (2008)

\bibitem {Grad}I. S. Gradshteyn and I. M. Ryzhik, Table of Integrals, Series,
and Products (Academic Press, New York 1979)
\end{thebibliography}
\end{document}